\documentclass[10pt]{article}
\usepackage[left=4cm,right=3cm,top=1.5cm,bottom=1cm,includeheadfoot]{geometry}
\usepackage{titlesec}
\usepackage[noblocks]{authblk}
\usepackage{fancyhdr}
\usepackage{graphicx}
\usepackage{lmodern}
\usepackage{csquotes}
\usepackage{amsmath}
\usepackage{amssymb}
\usepackage{bbold}
\usepackage{multicol}
\usepackage{lastpage} % for LastPage ref
\usepackage[dvipsnames]{xcolor}
\usepackage{xr-hyper}
\usepackage[hidelinks]{hyperref}
\usepackage[sort&compress]{cleveref}
\usepackage{orcidlink} % must come after hyperref for options to work

\usepackage[
    style=phys,
    biblabel=brackets,
    articletitle=true,
    url=false,
    eprint=false,
    doi=false,
    isbn=false,
    maxbibnames=4
]{biblatex}
\AtEveryBibitem{\clearfield{note}}

\titleformat{\section}{\large\bfseries}{\thesection)}{1.4em}{}
\titleformat{\subsection}{\normalsize\bfseries}{\thesubsection)}{0.95em}{}
\titleformat{\subsubsection}{\normalsize}{\thesubsubsection)}{0.55em}{}

\crefname{equation}{eq.\!}{eqs.\!}
\Crefname{equation}{Eq.\!}{Eqs.\!}
\crefname{figure}{fig.\!}{figs.\!}
\Crefname{figure}{Fig.\!}{Figs.\!}

\renewcommand{\vec}[1]{\boldsymbol{#1}} % spatial vectors

\newcommand*\diff[1][]{\mathop{}\!\mathrm{d}{#1}} % for differentials
\newcommand*\ddiff[2]{\frac{\diff[#1]}{\diff[#2]}} % for differentials
\newcommand{\dd}{\mathrm{d}^{d}}

\newcommand{\infint}{\int_{-\infty}^{\infty}}
\newcommand{\feq}{f^\mathrm{eq}}
\newcommand{\feqt}{\tilde{f}^\mathrm{eq}}

\newcommand{\cs}{c_\mathrm{s}}
\newcommand{\ia}{\alpha}
\newcommand{\ib}{\beta}
\newcommand{\ic}{\gamma}
\newcommand{\id}{\varepsilon}
\newcommand{\ie}{\sigma}

\newcommand{\an}[1]{a^{(#1)}}
\newcommand{\Hn}[1]{H^{(#1)}}

\newcommand{\x}{\vec{x}}
\renewcommand{\v}{\vec{v}}
\newcommand{\w}{\vec{s}}
\renewcommand{\c}{\vec{c}}
\renewcommand{\u}{\vec{u}}

\newcommand{\deltaf}{\Delta f}

\usepackage{wasysym}
\title{Lattice Boltzmann Methods with Anisotropic Equilibrium Distributions}
\newcommand{\shorttitle}{Anisotropic LBM}
\date{}

\author[1,2,*]{\orcidlink{0000-0001-9791-2724}\,Benjamin Kellers}
\author[1,2]{\orcidlink{0000-0002-5380-6791}\,Julius Weinmiller}
\author[1,2,3]{\orcidlink{0000-0003-1449-8172}\,Arnulf Latz}
\author[1,2]{\orcidlink{0000-0003-2336-6059}\,Timo Danner}

\affil[1]{%
German Aerospace Center (DLR), Institute of Engineering Thermodynamics,
Wilhelm-Runge-Str.\,10,
89081 Ulm, Germany}
\affil[2]{%
Helmholtz Institute Ulm Electrochemical Energy Storage (HIU),
Helmholtzstr.\,11,
89081 Ulm, Germany}
\affil[3]{%
Ulm University,
Institute of Electrochemistry,
Albert-Einstein-Allee~47, 
89081 Ulm, Germany}
\begin{document}

\maketitle
\renewcommand{\thefootnote}{\fnsymbol{footnote}}
\footnotetext[1]{Corresponding authors:
\href{mailto:benjamin.kellers@dlr.de}{benjamin.kellers@dlr.de}}
\renewcommand{\thefootnote}{\arabic{footnote}}

\thispagestyle{empty}
\pagestyle{fancy}

\fancyhead{} % clear all header fields
\fancyhead[L]{\textit{\shorttitle}}
\fancyhead[R]{\thepage{} of \pageref*{LastPage}}
\fancyfoot{} % clear all footer fields

\begin{abstract}
    \noindent Lattice Boltzmann methods are usually derived under the assumption of isotropy. In this work, we present a derivation of a Lattice Boltzmann method for anisotropic fluid flow. Starting from an anisotropic equilibrium distribution, we show a full derivation of the resulting lattice Boltzmann method. We ensure that our method correctly reproduces macroscopic behavior via Chapman-Enskog analysis for a single-relaxation time collision operator. As a result, we are able to show that a properly discretized anisotropic Maxwell-Boltzmann equilibrium does macroscopically in fact lead to an anisotropic variation of the Navier-Stokes equations. All desired properties of lattice Boltzmann methods, such as locality of the collision operator, isotropic discrete position and velocity space, or mass and momentum conservation are retained. While it is explicitly shown in the context of fluid flow, the presented scheme is straight-forward to adopt  to advection-diffusion problems.
\end{abstract}

\section{Introduction}
Isotropy is a central assumption in the derivation of classical lattice Boltzmann methods \cite{Krueger_2017:LBM_Book}. Not only do the position and velocity spaces have to be isotropic, the Maxwell-Boltzmann equilibrium distribution itself is isotropic. 

Anisotropic fluid flow, however, is not uncommon in the rheology of complex fluids. One can think of polymer emulsions \cite{Strautins_2007}, often containing 1D or 2D components \cite{Niu_2022}, or liquid crystals \cite{Nishikawa_2022}, where the anisotropy stems from the molecular shape itself. 
%Another setting are dissolved particles interacting with fields, e.g.\ iron particles in a magnetic field \todo{\cite{?}}. 
The hydrodynamics  for molecular liquids with anisotropic molecules is also relevant for  the interpretation of light scattering experiments \cite{Latz_2001,Franosch_2003}.
Another application is anisotropic advection-diffusion, which is a more common phenomenon since the diffusivity of many materials has a directional dependence.
Therefore, anisotropic heat and mass transport can intrinsically not be described by standard theory. This limits the application of the method in the relevant fields of anisotropic heat conduction or diffusion \cite{Wegener_2022}. 
%Additionally. flow behaviour as observed in polymer emulsions or liquid crystals cannot be captured using classical LBM schemes.

A very important application of high technical relevance is flow in anisotropic porous media. The pore structure resulting for example from irregular particle shapes  can introduce significant anisotropy in the resulting macroscopic flow field. In homogenized models this effective anisotropic flow behavior has to be treated accordingly to describe macroscopic flow patterns. 
% There have been attempts to model this through an anisotropic force field \cite{Seta_2009}. 

Other works related to anisotropic lattice Boltzmann models are often restricted by a very specific application, for example crystal growth \cite{Xing_2026}. 
An existing general approach \cite{Ginzburg_2013} is limited to advection-diffusion problems.
Moreover, multi-relaxation-time (MRT) schemes \cite{Chai_2016} have been presented in the literature, which can be used to create anisotropy in the resulting macroscopic equations \cite{Chai_2023}. MRT schemes introduce different relaxation time scales for the macroscopic moments resulting in enough free parameters to relax isotropy assumptions. We are not aware of a framework, that introduces anisotropy generically through the equilibrium distribution without the need to specify a collision scheme. Not only is it very natural to introduce the anisotropy through physical properties of the particle distributions, deriving a consistent model from first principles. In this way it is also applicable to a wide range of collision schemes and straight-forward to add to existing code bases.
The presented work addresses this gap by deriving a universal lattice Boltzmann method for anisotropic equilibrium distributions. We fully derive the discretization in this new setting and compute the resulting macroscopic laws through Chapman-Enskog theory for a single-relaxation time collision operator. Consequently, we arrive at anisotropic Navier-Stokes equations. The methodology is described in detail allowing transfer of the scheme from flow to other applications such as advection-diffusion problems or heat transport.
This is an important methodological development and provides new possibilities for the simulation of complex material behavior. 

This article is structured as follows: In \Cref{sec:methods} we first give a short introduction to the Boltzmann equation, before introducing the anisotropic modification of the equilibrium distribution and performing the discretization of the velocity space. 
In \Cref{sec:CET} we perform the Chapman-Enskog analysis via perturbation in orders of Knudsen numbers yielding the governing macroscopic equations. 
As a numerical example we show the change of decay frequencies as a result of anisotropy for Taylor-Green vortices in \Cref{sec:TG}.
We end with a short conclusion and an outlook on possible future extensions in \Cref{sec:conclusion}. For ease of reading, most of the required mathematical background is given in the Appendix.

\section{Methods}
\label{sec:methods}

\subsection{The Boltzmann Equation}

Lattice-Boltzmann methods (LBM) aim to numerically solve the Boltzmann equation
\begin{equation}
    \ddiff{f}{t} = \Omega,
\end{equation}
which expresses the time evolution of a particle distribution $f$ under the influence of interparticle collisions described by an operator $\Omega$. In the most general sense a particle distribution function $f$ maps the density distribution of a large number of non-interacting non-relativistic particles as a function of the microscopic particle velocities $\v$, time $t$, and spatial position $\x$. The distribution is normalized such that the zeroth order moment yields the mass density $\rho$ and the first order moment the momentum density $\rho u_\ia$ as time-dependent fields, i.e.\ 
\begin{equation}
    \infint f(t,\x,\v) \dd v = \rho(t,\x),
\label{eq:A1}
\end{equation}
\begin{equation}
    \infint f(t,\x,\v) v_\ia \dd v = \rho(t,\x) u_\ia(t,\x).
\label{eq:A2}
\end{equation}
Throughout this work, Greek indices run over the $d$ spatial or velocity dimensions. Given enough time, inter-particle collisions will drive the system towards a local equilibrium state described by an equilibrium distribution $\feq$ which depends only on the first two macroscopic moments and the microscopic particle velocities. The distribution will also yield the same moments to first order, i.e.\ 
\begin{equation}
    \infint \feq(\rho,\u,\v) \dd v = \rho(t,\x),
\label{eq:A3}
\end{equation}
\begin{equation}
    \infint \feq(\rho,\u,\v) v_\ia \dd v = \rho(t,\x) u_\ia(t,\x).
\label{eq:A4}
\end{equation}
% These must hold, otherwise we will have a hard time constructing collision operators that conserve mass and momentum. And this would violate our physical intuition of a picture in which perfectly elastic collisions among a fixed number of fluid particles occur.

In local equilibrium, the space-time dependence enters only implicitly through the macroscopic fields. Here, and in the following, we only consider an isothermal system, where we can disregard the second order moment related to energy.

Usually, it is assumed that the interparticle collisions have no preferred direction and will hence even out any angular dependence of an initial state very rapidly, resulting in a isotropic and isothermal equilibrium described by a Maxwell-Boltzmann-type distribution
\begin{equation}
\bar{f}^\mathrm{eq}(\rho,\u,\v) = \rho(t,\x) \frac{1}{\sqrt{2\pi}^d} 
        \exp\left(-\frac{[v_\ia - u_\ia(t,\x)]\delta_{\ia\ib}[v_\ib - u_\ib(t,\x)]}{2} \right),
\end{equation}
where $\delta_{\ia\ib}$ is the Kronecker delta. Since we work in the isothermal model, we choose units in which the temperature constant is unity to simplify notation. This expression utilizes Einstein's summation convention where we sum over repeated indices, which is applied throughout this work.

In the following, we derive an anisotropic LBM by skewing this distribution. We will develop the whole discretization framework based on this assumption and compute the governing macroscopic equations of a system described by this type of equilibrium.

\subsection{Expansion of an Anisotropic Equilibrium Distribution}

One of the simplest forms of a Maxwell-Boltzmann-like isothermal equilibrium distribution which introduces anisotropy can in suitable units and $d$ dimensions be written as
\begin{equation}
    \feq(\rho,\u,\v) = \rho(t,\x) \frac{1}{\sqrt{2\pi}^d} 
    \exp \left(-\frac{[v_\ia-u_\ia(t,\x))] A_{\ia\ib} [v_\ib-u_\ib(t,\x)]}{2}\right).
\label{eq:equil}
\end{equation}
Since we limit our investigation to the case where $\feq$ remains Boltzmann-like, the expression in the exponential remains a quadratic form. Any quadratic form can be described by a unique symmetric matrix, which we call $A$. To not pick up extra normalization factors in all expressions, we additionally require $\det A=1$. Physically, $A$ describes the strength of the directional dependence of the equilibrium velocity distribution.

% Das Folgende ist nur wichtig, wenn man den Formalismus über eine Koordinatentransformation beschreiben möchte.
%We now further constrain $A$ by requiring $v_\ia A_{\ia\ib} v_\ib > 0 \quad\forall\; \v \in \mathbb{R}^d \setminus \{\vec{0}\}$. $A$ is then called positive definite which comes with special properties, one of which we care about in particular: If $A$ is positive definite and symmetric, there exist many matrices $C$ such that $A_{\ia\ic} = C_{\ia\ib} C_{\ib\ic}$ or $A_{\ia\ic} = C_{\ib\ia} C_{\ib\ic}$. But there exists exactly one matrix $B$ that is also symmetric and positive definite, such that $A_{\ia\ic}=B_{\ia\ib} B_{\ib\ic} = B_{\ib\ia}B_{\ib\ic}$. We call $B_{\ia\ib}$ the square root of $A_{\ia\ib}$, often written as $A_{\ia\ib}^{1/2}$.

The macroscopic moments of this new anisotropic equilibrium are still
\begin{equation}
    \infint \feq \dd{v} = \rho(t,\x), 
\end{equation}
\begin{equation}
    \infint \feq v_\ia \dd{v} = \rho(t,\x) u_\ia(t,\x),
\end{equation}
cf.\ \Cref{app:integrals}. 
% If we define $\u$ as in \Cref{eq:equil} with $A$ acting on it, the first two moments do not include a correction term, since $\u$ already carries it implicitly. This becomes apparent when solving the integral by hand through a change of variables.
The explicit treatment of a full exponential makes the theory laborious and any simulations computationally expensive.
Since we only enforce to reproduce the first two moments, our physical observables, correctly, it is sufficient to approximate the equilibrium distribution.
In this work, we limit the discussion to the single-relaxation time collision operator (BGK) \cite{BGK,BGK2,BGK3}, 
which recovers macroscopically the Navier-Stokes equations (cf.\ \Cref{app:navier_stokes}) using an expansion up to second order in velocity for the isotropic case. An anisotropic equilibrium, however, benefits from third order expansion, as will we shown in \Cref{sec:CET}. Using only second order, the anisotropic Navier-Stokes equations will include a discretization error depending on the anisotropy, see also \Cref{app:perturbation_moment}.

To expand the exponential, LBM utilizes Hermite polynomial expansion because the resulting distribution has suitable properties for the following velocity space discretization. Details can be found in \Cref{app:Hermite} and Ref.\ \cite{Krueger_2017:LBM_Book}. Expanded to third order, the equilibrium distribution can be approximated as
\begin{equation}
    \feq(\v) \simeq \omega(\v) \sum_{n=0}^3 \frac{1}{n!} 
    \langle \an{n}(\feq) , \Hn{n} \rangle,
\end{equation}
where $\omega$ is the generator of the Hermite polynomials $\Hn{n}$, cf.\ \Cref{app:Hermite}. The operation $\langle\cdot,\cdot\rangle$ is the full contraction of the $n$-rank tensors $\Hn{n}$ and $\an{n}$.
The expansion coefficients are projections onto the Hermite polynomials, i.e.\ 
\begin{equation}
    \an{n}(\feq) = \infint \feq(\v) \Hn{n}(\v) \dd v,
\label{eq:expcoeff}
\end{equation}
which, to third order, results in
\begin{equation}
\begin{split}
    \feqt &(t,\x,\v) = \omega(\v) \sum_{n=0}^3 \frac{1}{n!} 
    \langle \an{n}(\feq) , \Hn{n} \rangle \\
    &\qquad\quad\; = \omega(\v) \rho(t,\x) \left[ 
        1
        + u_\ia v_\ia
        + \frac{1}{2} (u_\ia u_\ib + K_{\ia\ib})(v_\ia v_\ib - \delta_{\ia\ib})
        \right. \\
    &\left.
        + \frac{1}{6}(
        u_\ia u_\ib u_\ic + K_{\ia\ib} u_\ic + K_{\ib\ic} u_\ia + K_{\ic\ia} u_\ib)
            \left( v_\ia v_\ib v_\ic 
                - v_\ia \delta_{\ib\ic} 
                - v_\ib \delta_{\ic\ia} 
                - v_\ic \delta_{\ia\ib} 
                \right)\right],
\end{split}
% \label{eq:3rd_order_equilibrium}
\end{equation}
where $u_\ia=u_\ia(t,\x)$, which we suppressed for the sake of readability. The matrix $K$ is given by
\begin{equation}
    K_{\ia\ib} = A^{-1}_{\ia\ib} - \delta_{\ia\ib}.
\label{eq:K-term}
\end{equation}
Calculation of the first to moments confirms that truncation does not change relevant macroscopic properties.
We find
\begin{equation}
    \infint \feqt \dd{v} 
    = \infint \feqt \Hn{0} \dd{v}
    = \an{0}(\feq)
    = \rho, 
\label{eq:14}
\end{equation}
\begin{equation}
    \infint \feqt v_\ia \dd{v} 
    = \infint \feqt \Hn{1}_\ia \dd{v}
    = \an{1}_\ia(\feq)
    = \rho u_\ia,
\label{eq:15}
\end{equation}
cf.\ \Cref{eq:expcoeff} and \Cref{eq:hermite_polynomials} in \Cref{app:Hermite}.
It is not surprising that the relevant moments remain unchanged, since the Hermite polynomials are orthogonal. To a certain degree, they separate orders of velocity. Taking these integrals corresponds to projecting out the expansion coefficients $\an{n}(\feq)$.

\subsection{Discretization of the Velocity Space}

For calculation of the macroscopic quantities \Cref{eq:14,eq:15} we still have to compute the full velocity integrals to get the macroscopic quantities as moments of the distribution function. Just as in isotropic LBM, we can convert velocity integrals into finite sums through Gaussian quadrature. Here, we construct a weighted sum of orthogonal polynomials that are evaluated at specific points, referred to as nodes or abscissae. There are numerous variants of such quadratures and the choice mostly depends on the range of the interval to be evaluated. For the range $(-\infty,\infty)$ the required method is Gauss-Hermite quadrature. It shares its generator $\omega$ with the Hermite polynomials and considerably simplifies the following treatment of the integrals. An overview of the method is available in \Cref{app:quadrature} or Ref.\ \cite{Krueger_2017:LBM_Book}.

In general, we encounter expressions including integrals of the form of \Cref{eq:expcoeff} for the macroscopic moments. Note that in the following the function $\feq$ can be considered the truncated one without affecting the accuracy of the scheme. Therefore, in the derivation below we refer to the truncated equilibrium distribution just as $\feq$, if not otherwise specified.

We now utilize the relation of the Hermite expansion and the Gauss-Hermite quadrature and split $\feq$ (\Cref{eq:3rd_order_equilibrium}) into three parts: the normalization $\rho$, the generator function $\omega$, and the remaining polynomial $\mathcal{P}$ which depends on $\v$ and $\u(t,\x)$. This yields
\begin{equation}
    \an{n}(\feq) = \infint \omega(\v)\rho(t,\x) \mathcal{P}(\v,\u) \Hn{n}(\v) \dd v.
\end{equation}
The quadrature procedure converts these integrals into sums by evaluating only a fixed set of points in the velocity space. Hence, the continuous velocity space essentially becomes discrete. The expansion coefficients are therefore given by
\begin{equation}
    \an{n}(\feq) = \sum_{i=1}^q w_i \rho(t,\x)\mathcal{P}(\v_i)\Hn{n}(\v_i),
\label{eq:quadrature}
\end{equation}
where $q$ depends on the desired accuracy. Another convenient property of the Gauss-Hermite quadrature in contrast to other variants is that the expression remains exact for suitably large $q$. The size of $q$ depends on the order of the product $\mathcal{P}\Hn{n}$.

To not stream off-lattice, spatial and velocity lattices need to overlap. But the nodes or abscissae, $\v_i$, do not agree with a spatial standard lattice using $\Delta x = 1$.  It is hence standard practice in LBM to rescale the velocities to 
\begin{equation}
    \v_i = \frac{\c_i}{\cs},\quad \u = \frac{\w}{\cs}.
\end{equation}
The constant factor $\cs$ could also be absorbed into the units. The updated equilibrium distribution reads
\begin{equation}
\begin{split}
    \feq (\c_i)
    = \omega(\c_i) \rho \left[
        1 \right.
        &+ \left.\frac{c_{i\ia} s_\ia}{\cs^2}
        + \frac{1}{2\cs^4} (s_\ia s_\ib + \cs^2 K_{\ia\ib})(c_{i\ia} c_{i\ib} - \cs^2\delta_{\ia\ib})
        \right. \\
        &+ \frac{1}{6\cs^6}
            \left(s_\ia s_\ib s_\ic + \cs^2 K_{\ia\ib} s_\ic + \cs^2 K_{\ib\ic} s_\ia + \cs^2 K_{\ic\ia} s_\ib \right)\\
        &\left.\times\left( c_{i\ia} c_{i\ib} c_{i\ic} 
            - c_{i\ia} \cs^2 \delta_{\ib\ic} 
            - c_{i\ib} \cs^2 \delta_{\ic\ia} 
            - c_{i\ic} \cs^2 \delta_{\ia\ib} 
            \right)
        \right].
\end{split}
\label{eq:3rd_order_equilibrium}
\end{equation}
Aiming for a clean notation, we define the discrete equilibrium function 
\begin{equation}
    \feq_i(t,\x) = \frac{w_i}{\omega(\c_i)} \feq(t,\x,\c_i),
\label{eq:feq_i}
\end{equation}
with the weights $w_i$ included and perform the quadrature, \Cref{eq:quadrature}, yielding the physically relevant macroscopic moments as sums in the discretized velocity space as
\begin{equation}
    \rho = \an{0}(\feq) = \sum_{i=1}^q \feq_i,
\label{eq:a0discrete}
\end{equation}
and 
\begin{equation}
    \rho u_\ia = \an{1}(\feq) = \sum_{i=1}^q \feq_i c_{i\ia}.
\label{eq:a1discrete}
\end{equation}

Similarly, we can discretize the particle distribution function $f$. Without repeating all the steps we just note that since we want to conserve mass and momentum, the distributions yield the same moments up to first order, cf.\ \Cref{eq:A1,eq:A2,eq:A3,eq:A4}. From this follows that the first two expansion coefficients must be identical, such that
\begin{equation}
    \rho = \an{0}(\feq) = \an{0}(f) =  \sum_{i=1}^q f_i,
\label{eq:a0discrete_f}
\end{equation}
and 
\begin{equation}
    \rho u_\ia = \an{1}(\feq) = \an{1}(f) = \sum_{i=1}^q f_i c_{i\ia},
\label{eq:a1discrete_f}
\end{equation}
with $f_i$ defined analogously to \Cref{eq:feq_i}.

Since the space-time discretization does not require modifications compared to the isotropic case, we will skip it here and refer the interested reader to Ref.\ \Cite{Krueger_2017:LBM_Book} instead.

\section{Chapman-Enskog Theory}
\label{sec:CET}
The Chapman-Enskog \cite{Chapman_Enskog} analysis will provide information about the macroscopic physical equations that are induced by the microscopic discrete formalism we developed above. A detailed but palatable derivation for the isotropic case can again be found in Ref.\ \Cite{Krueger_2017:LBM_Book}. We will follow a similar derivation in the procedure below.

We want to derive the behavior of the anisotropic equilibrium function under single-relaxation time collision (BGK)
\cite{BGK,BGK2,BGK3}, 
corresponding to the fully discrete Boltzmann equation
\begin{equation}
    f_i(t+\Delta t,x+\c_i \Delta t) - f_i(t,\x)
    = \frac{\Delta t}{\tau} [\feq_i(t,\x) - f_i(t,\x)],
\label{eq:LBE_BGK}
\end{equation}
where $\tau$ is the relaxation time. Here, we performed the time discretization to first order only, since it is equivalent to second order assuming a suitable redefinition of $f_i$, compare Appendix A.5 in Ref.~\cite{Krueger_2017:LBM_Book}.%
\footnote{%
For BGK, it can be shown that there exists a distribution $\bar{f}_i$ with the same moments as $f_i$ and a corresponding relaxation time $\bar{\tau} = \tau + \Delta t /2$, with which the second order time accurate scheme has the same form as the first order \Cref{eq:LBE_BGK}.
} 
Some details about the expected form of the resulting Navier-Stokes equations for single-relaxation time collisions and their anisotropic modification are given in \Cref{app:navier_stokes}.

Performing the analysis on this collision operator, we first note that a general discrete distribution $f_i$ deviates from the equilibrium by some amount $\Delta f_i$, i.e.\ 
\begin{equation}
    f_i = \feq_i + \deltaf_i 
    \quad \Leftrightarrow \quad  
    \Delta f_i = f_i - \feq_i.
\label{eq:deltaf}
\end{equation}
This can be understood as the collection of particles consisting of an equilibrated part plus a (small) non-equilibrium perturbation. The macroscopic dynamic properties of the fluid fully depend on this perturbation. Similar to other perturbation theories, the perturbation can be expanded in orders of some ordering-parameter, in our case the Knudsen number 
\begin{equation}
    \mathrm{Kn} = \frac{\lambda}{L},
\end{equation}
with $\lambda$ and $L$ being the mean free path of fluid particles and the characteristic physical length scale of the problem, respectively.

We expand the distribution as
\begin{equation}
    f_i = \sum_{n=0}^\infty \epsilon^n f_i^{(n)} 
    = f_i^{(0)} + \sum_{n=1}^\infty \epsilon^n f_i^{(n)} 
    = \feq_i + \sum_{n=1}^\infty \epsilon^n f_i^{(n)} 
    \stackrel{(\ref{eq:deltaf})}{=} \feq_i + \Delta f_i,
\label{eq:perturbation_expansion}
\end{equation}
where powers of expansion coefficients $\epsilon$ determine the order in Kn, i.e.\ 
\begin{equation}
    \frac{f^{(n)}}{\feq} = \mathcal{O}(\mathrm{Kn}^n).
\end{equation}

By combining \Cref{eq:a0discrete,eq:a1discrete,eq:a0discrete_f,eq:a1discrete_f} it follows that
\begin{equation}
    \sum_i \feq_i = \sum_i f_i \quad \stackrel{(\ref{eq:deltaf})}{\Longrightarrow} \quad \sum_i \Delta f_i = 0
\label{eq:Dfzero0}
\end{equation}
and
\begin{equation}
    \sum_i \feq_i \c_i = \sum_i f_i \c_i \quad \stackrel{(\ref{eq:deltaf})}{\Longrightarrow} \quad \sum_i \Delta f_i \c_i = \vec{0}.
\label{eq:Dfzero1}
\end{equation}
The expansion in \Cref{eq:perturbation_expansion} makes $\Delta f_i$ a polynomial in $\epsilon$. By comparison of coefficients \Cref{eq:Dfzero0,eq:Dfzero1} imply even stricter relations, namely
\begin{equation}
    \sum_i f_i^{(n)} = 0 \quad\forall n \geq 1,
\label{eq:prop1}
\end{equation}
and
\begin{equation}
    \sum_i f_i^{(n)} \c_i = \vec{0} \quad\forall n \geq 1.
\label{eq:prop2}
\end{equation}
Next, we expand the whole left hand side of \Cref{eq:LBE_BGK} into a Taylor series up to second order, cf.\ \Cref{eq:Taylor_expansion} in \Cref{app:Taylor_expansion}, and apply \Cref{eq:deltaf} to the right hand side, which yields
\begin{equation}
    \Delta t(\partial_t + c_{i\ia}\partial_\ia) f_i
    + \frac{\Delta t^2}{2}(\partial_t + c_{i\ia}\partial_\ia)^2 f_i 
    + \mathcal{O}(\Delta t^3)
    = -\frac{\Delta t}{\tau}\Delta f_i.
\label{eq:step1}
\end{equation}
The treatment of a second derivative would be complicated. To eliminate it, we apply the operator $\frac{\Delta t}{2}(\partial_t + c_{i\ia}\partial_\ia)$ to both sides of \Cref{eq:step1}, i.e.\  
\begin{equation}
    \frac{\Delta t^2}{2}(\partial_t + c_{i\ia}\partial_\ia)^2 f_i 
    + \mathcal{O}(\Delta t^3)
    = -\frac{\Delta t^2}{2\tau}(\partial_t + c_{i\ia}\partial_\ia) \Delta f_i,
\label{eq:step2}
\end{equation}
and subtract the resulting \Cref{eq:step2} from \Cref{eq:step1}, yielding
\begin{equation}
    \Delta t(\partial_t + c_{i\ia}\partial_\ia) f_i
    = -\frac{\Delta t}{\tau}\Delta f_i 
    + \frac{\Delta t^2}{2\tau}(\partial_t + c_{i\ia}\partial_\ia) \Delta f_i,
\label{eq:step3}
\end{equation}
where we neglected the $\mathcal{O}(\Delta t^3)$ terms to simplify the expression.

The idea is to get a set of multiple equations, each corresponding to the next correction when considering higher and higher orders of Kn. 
The relevant order for the Navier-Stokes-regime is $\mathcal{O}(\mathrm{Kn}^2)$.\cite{Krueger_2017:LBM_Book} Hence, we expand the time derivative in a similar fashion as $\Delta f_i$ in \Cref{eq:perturbation_expansion} up to second order, i.e.\ 
\begin{equation}
    \partial_t f_i = (\epsilon \partial_t^{(1)} + \epsilon^2 \partial_t^{(2)})f_i.
\end{equation}
The spatial derivative is sufficient at first order, but we want to introduce the corresponding notation
\begin{equation}
    c_{i\ia} \partial_\ia f_i = \epsilon c_{i\ia} \partial_\ia^{(1)}f_i.
\end{equation}
Applying these definitions to \Cref{eq:step3} yields
\begin{equation}
    \Delta t \left[
        \left( 
            \epsilon \partial_t^{(1)} + \epsilon^2 \partial_t^{(2)} 
        \right)
        + \epsilon c_{i\ia} \partial_\ia^{(1)}
    \right] f_i
    = -\frac{\Delta t}{\tau}\Delta f_i 
    + \frac{\Delta t^2}{2\tau}
    \left[
        \left(
            \epsilon \partial_t^{(1)} + \epsilon^2 \partial_t^{(2)}
        \right)
        + \epsilon c_{i\ia} \partial_\ia^{(1)}
    \right] \Delta f_i.
\end{equation}
Using \Cref{eq:perturbation_expansion} we expand $\Delta f_i$ and $f_i$ to second order and split the result into two equations, one for each order of $\epsilon$, yielding
\begin{align}
    % \mathcal{O}(\epsilon)&: \qquad
    \epsilon
    \left( \partial_t^{(1)} + c_{i\ia} \partial_\ia^{(1)} \right) \feq_i
    &= -
    \frac{\epsilon}{\tau} f_i^{(1)}, 
\label{eq:CE1} 
    \\
    % \mathcal{O}(\epsilon^2)&: \qquad
    \epsilon^2
    \partial_t^{(2)} \feq_i
    + 
    \epsilon^2
    \left(1 - \frac{\Delta t}{2\tau} \right) 
        \left( \partial_t^{(1)} + c_{i\ia}\partial_\ia^{(1)} \right) f_i^{(1)}
    &= -
    \frac{\epsilon^2}{\tau} f_i^{(2)}.
\label{eq:CE2}
\end{align}
Again, the macroscopic behavior is determined by the first two moments, zeroth and first order. And the second order moment is essentially a correction term. From the Gauss-Hermite quadrature we know that taking moments in the discrete velocity space corresponds to multiplying expressions by power of $\c_i$ before summing over $i$. We can take moments of the full equations by applying the operations to both sides. Considering \Cref{eq:prop1,eq:prop2}, the zeroth moments of \Cref{eq:CE1,eq:CE2} are
\begin{equation}
    \epsilon
    \partial_t^{(1)} \rho 
    + \epsilon \partial_\ia^{(1)}(\rho u_\ia) = 0,
\label{eq:1st_order_mass}
\end{equation}
and
\begin{equation}
    \epsilon^2
    \partial_t^{(2)} \rho = 0,
\end{equation}
respectively. Adding both equations and reversing the derivative expansions yields the governing equation for the zeroth order moment (mass), i.e.\ the continuity equation
\begin{equation}
    \partial_t \rho 
    + \partial_\ia(\rho u_\ia) = 0.
\end{equation}

With moment tensors $M_{\ia_1...\ia_n}$ defined analogously to \Cref{app:moments}, the first order moments of \Cref{eq:CE1,eq:CE2} are
\begin{equation}
    \epsilon \partial_t^{(1)}(\rho u_\ia)
    + \epsilon \partial_\ib^{(1)} M_{\ia\ib}(\feq_i) = 0
\label{eq:1st_order_momentum}
\end{equation}
and
\begin{equation}
    \epsilon^2 \partial_t^{(2)}(\rho u_\ia)
    + \epsilon^2  \partial_\ib^{(1)} \left( 1 - \frac{\Delta t}{2\tau} \right) M_{\ia\ib}(f_i^{(1)}) = 0,
\end{equation}
respectively. Again, adding these and recombining the expansions yields the governing equation for the first order moment (momentum), i.e.\ 
\begin{equation}
    \partial_t(\rho u_\ia) + \partial_\ib M_{\ia\ib} (\feq_i) = -\epsilon 
    \partial_\ib \left[ 
        \left( 1 - \frac{\Delta t}{2\tau} \right) 
        M_{\ia\ib}(f^{(1)}_i)
    \right].
\label{eq:CE3}
\end{equation}
The higher order moments of the equilibrium encountered in this equation can be evaluated according to \Cref{app:moments}. But up to this point the second order moment of the perturbation $f^{(1)}_i$ is unknown. However, taking the second order moment of \Cref{eq:CE1} results in a defining equation for it, i.e.\ 
\begin{equation}
    \epsilon \partial_t^{(1)} M_{\ia\ib}(\feq_i) 
    + \epsilon\partial_\ic^{(1)} M_{\ia\ib\ic}(\feq_i)
    = -\frac{\epsilon}{\tau}  M_{\ia\ib}(f_i^{(1)}).
\label{eq:CE4}
\end{equation}
This equation provides us with the unknown moment in terms of second and third order moments of the equilibrium, which we can compute. Corresponding calculations can be found in \Cref{app:perturbation_moment}. The needed equilibrium moments can again be found in \Cref{app:moments}. Using \Cref{eq:CE4}, we find the perturbation moment
\begin{equation}
    M_{\ia\ib}(f_i^{(1)}) = 
    - \tau \rho \cs^2 \left(
          A^{-1}_{\ib\ic} \partial_\ic^{(1)} s_\ia
        + A^{-1}_{\ia\ic} \partial_\ic^{(1)} s_\ib
    \right).
\end{equation}
Inserting the second order moment of the perturbation into \Cref{eq:CE3} we arrive at
\begin{equation}
    \partial_t(\rho u_\ia) + \partial_\ib M_{\ia\ib} (\feq_i) = \epsilon 
    \partial_\ib \left[
        \rho \cs^2 \left( \tau - \frac{\Delta t}{2} \right) 
            \left( 
            A^{-1}_{\ib\ic} \partial_\ic^{(1)} s_\ia
            + A^{-1}_{\ia\ic} \partial_\ic^{(1)} s_\ib
        \right)
    \right].
\end{equation}
After inserting the expression for the equilibrium moment and reversing the derivative expansion we finally arrive at
\begin{equation}
    \partial_t(\rho s_\ia) 
    + \partial_\ib \left( 
          \rho \cs^2 A^{-1}_{\ia\ib}
        + \rho s_\ia s_\ib
    \right) 
    = \partial_\ib \left[ 
        \rho \cs^2 \left( \tau - \frac{\Delta t}{2} \right) 
            \left(
                A^{-1}_{\ib\ic} \partial_\ic s_\ia
                + A^{-1}_{\ia\ic} \partial_\ic s_\ib
            \right)
        \right].
\label{eq:NSprefinal}
\end{equation}
By defining a shear viscosity
\begin{equation}
    \eta = \rho \cs^2 \left( \tau - \frac{\Delta t}{2} \right) 
\end{equation}
and applying the equation of state
\begin{equation}
    p = \rho \cs^2,
\end{equation}
where $p$ is the pressure of the fluid, we are left with a more familiar form of \Cref{eq:NSprefinal}, i.e.\ 
\begin{equation}
    \partial_t(\rho s_\ia) 
    + \partial_\ib ( \rho s_\ia s_\ib ) 
    = 
    - A^{-1}_{\ia\ib} \partial_\ib p
    + \partial_\ib \left[ 
        \eta
            \left(
                A^{-1}_{\ib\ic} \partial_\ic s_\ia
                + A^{-1}_{\ia\ic} \partial_\ic s_\ib
            \right)
        \right].
\label{eq:NSprefinal2}
\end{equation}
With our derivation, we imply that the bulk viscosity 
\begin{equation}
    \zeta = \frac{2}{3} \eta,
\end{equation}
hence, two terms otherwise present in the Navier-Stokes equations cancel for us.

\Cref{eq:NSprefinal2} is only in this form because we expanded the equilibrium to third order in velocity. Otherwise a $K$-dependent discretization error would remain. In the next section we will demonstrate that the third order expansion is also strictly necessary when capturing phenomena that are related to higher order velocity excitations.

The reader might also have expected an anisotropic transformation acting on the macroscopic velocity $\w$. This is in fact the case, although not explicitly visible. By defining the equilibrium as we have in \Cref{eq:equil}, $A$ is already applied to $\w$, hence $\w$ is already anisotropically skewed by definition and requires no additional transformation.

\section{Taylor-Green-Vortex Decay}
\label{sec:TG}

To demonstrate the behavior and properties of the anisotropic BGK model, we investigate the decay of a Taylor-Green vortex {\cite{TaylorGreen_1937}}. It is a common example to measure the time accuracy of fluid flow solvers, since it has an analytical time-dependent solution. For us the vortex itself is interesting since it contains higher order velocity terms, which are necessary to highlight the importance of the third order expansion of the equilibrium distribution, {\Cref{eq:3rd_order_equilibrium}}.
In two spatial dimensions using coordinates $\x = (x,y)$, the problem can be stated as follows {\cite{Krueger_2017:LBM_Book}}: A vortex-velocity field given by
\begin{equation}
    \vec{u}(\x) = u_\mathrm{max} 
    \begin{pmatrix}
        -\sqrt{k_y/k_x} \cos(k_x x)\sin(k_y y)\\
        \phantom{-}\sqrt{k_x/k_y}  \cos(k_y y)\sin(k_x x)
    \end{pmatrix},
\end{equation}
will on a fully periodic domain of size $l_x\times l_y$ decay exponentially with some characteristic decay frequency $\omega_\mathrm{d}$. Here, $k_\ia$ are wavenumbers, i.e.\ $k_\ia = 2\pi / l_\ia$, and $u_\mathrm{max}$ is the maximal velocity magnitude. The decay of such a vortex field over time $t$ is hence given by
\begin{equation}
    \vec{u}(t,\x) = u_\mathrm{max} 
    \begin{pmatrix}
        -\sqrt{k_y/k_x} \cos(k_x x)\sin(k_y y)\\
        \phantom{-}\sqrt{k_x/k_y}  \cos(k_y y)\sin(k_x x)
    \end{pmatrix}
    \exp(-\omega_\mathrm{d}t).
\label{eq:TG_u}
\end{equation}

The pressure field will be
\begin{equation}
    p(t,\x) = p_0 - \frac{\rho u_\mathrm{max}^2}{4}
        \left[
        \frac{k_y}{k_x}\cos(2k_x x) + \frac{k_x}{k_y}\cos(2k_y y)
        \right]
        \exp(-2\omega_\mathrm{d} t).
\label{eq:TG_p}
\end{equation}

We will use a quadratic simulation domain, $l_x=l_y$. Usually, one investigates cross terms, for example in the stress tensor $\sigma$, by simulating a domain with $l_x\neq l_y$, since otherwise the stress tensor cross terms will be zero, i.e.\ $\sigma_{xy} = \sigma_{yx} = 0$. With our anisotropic model, however, we will see the effects of the velocity cross terms in the equilibrium distribution even in quadratic domains.

For our setup, we chose to simulate the vortex decay over 400 time steps on a $l_x\times l_y = 64\times 64$ domain. The BGK relaxation time $\tau$ is unity and $u_\mathrm{max}=0.02$. In isotropic LBM, these parameters correspond to a kinematic viscosity $\nu = (2\tau-1)/6$. The analytical solution for the Taylor-Green vortex predicts in this case a decay frequency of 
\begin{equation}
    \omega_\mathrm{d} = \nu (k_x^2 + k_y^2) 
        = \frac{2}{6} \left(\frac{2\pi}{64}\right)^2
        = 3.21\times10^{-3}
\end{equation}
in lattice units.

To describe the anisotropy, we define a matrix
\begin{equation}
    \tilde{A} = \begin{pmatrix}
        \tilde{A}_{xx} & \tilde{A}_{xy} \\
        \tilde{A}_{yx} & \tilde{A}_{yy},
    \end{pmatrix}
\end{equation}
which is proportional to $A$ in {\Cref{eq:equil}}, but not yet normalized, meaning that the determinant is not necessarily unity. The matrix $A$ is then simply
\begin{equation}
    A = \frac{1}{\sqrt{\det \tilde{A}}} \tilde{A}.
\end{equation}

%%%%%%%%%%%%%%%%%%%%%%%%%%%%%%%%%%%%%%%%%%%%%%%%%%%%%%%%%%%%%%%%%%%%%%%%%%%%%%%
\begin{figure}
    \centering
    \includegraphics[width=0.99\linewidth]{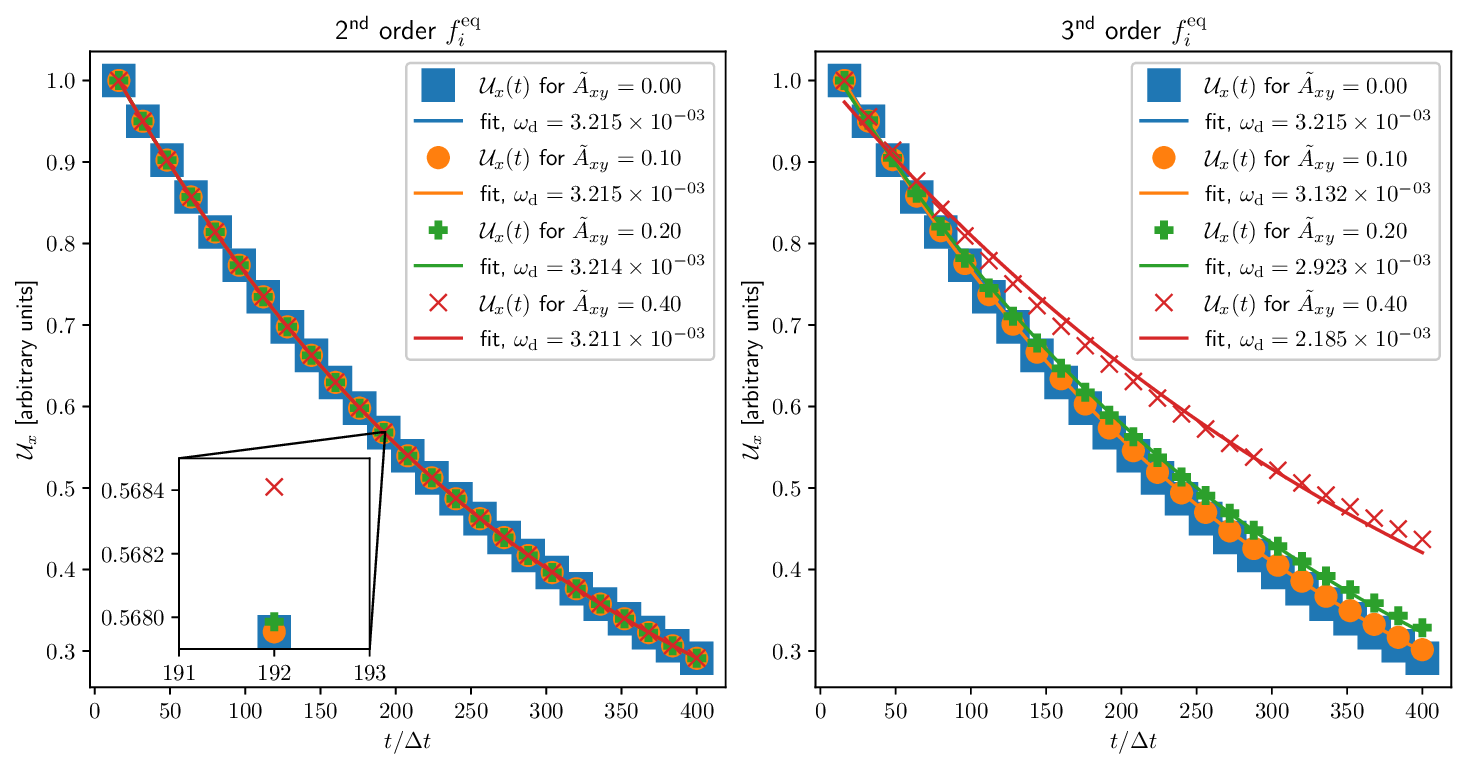}
    \caption{Exponential decay of the Taylor-Green vortex. Velocity U$_x$ as function of simulation time. The velocity is scaled to the highest value at t=0, because it is a sum over the whole domain with an arbitrary numerical value which is captured by the prefactor $a$ in the fit function $\mathcal{F}(t) = a\exp(-\omega_\mathrm{d}t)$. Left: Second order approximation of the equilibrium function. A zoom into the results using the second order approximation shows that the values vary in the fourth decimal only. Even the expected order from small to large values of $A_{xy}$ is produced (compare with the zoomed inset). Right: Third order approximation of the equilibrium function. Velocity data splits up significantly and we predict significantly different decay frequencies. }
    \label{fig:TG_decay}
\end{figure}
%%%%%%%%%%%%%%%%%%%%%%%%%%%%%%%%%%%%%%%%%%%%%%%%%%%%%%%%%%%%%%%%%%%%%%%%%%%%%%%

Since we want to focus on the cross terms, we leave the diagonals as unity for all simulations, i.e.\ $\tilde{A}_{xx} = \tilde{A}_{yy} = 1$. For $\tilde{A}_{xy} = \tilde{A}_{yx}$ we chose four different values, $\tilde{A}_{xy} = 0.0,\,0.1,\,0.2,\,0.4 $, such that the first case corresponds to an isotropic model with $\tilde{A} = A = \mathbb{1}$. We will compare the solutions of the decay frequencies for simulations using the third order approximations of the equilibrium distributions ( {\Cref{eq:3rd_order_equilibrium}}) with solutions simulated using equilibrium functions up to approximated second order only. The domain velocity is initialized with the Taylor-Green velocity field, {\Cref{eq:TG_u}}, at $t=0$. The density depends on the pressure field, {\Cref{eq:TG_p}}, and is set to $\rho(\x) = 1 + 3 p(t=0,\x)$. 

As measure for the decay of the velocity field we calculate the sum of absolute x-components of the velocity vector over the whole domain, i.e.\ 
\begin{equation}
    \mathcal{U}_x(t) = \sum_\mathrm{\{\x\}} u_x(t,\x),
\end{equation}
for 25 equally spaced time intervals. Since both components decay with the same frequency in a quadratic domain, \Cref{eq:TG_u}, one of them is sufficient to determine $\omega_\mathrm{d}$. 

The results are presented in {\Cref{fig:TG_decay}}. The left hand side of the figure gives simulation results using the second order accurate approximation of the equilibrium function. Data points at time intervals were fitted with \texttt{optimize.curve\_fit} from the Python library SciPy {\cite{SciPy}} using the fit function $\mathcal{F}(t) = a\exp(-\omega_\mathrm{d}t)$. Simulations recover the characteristic decay frequency for the isotropic case. However, the second order approximation is not sufficient to simulate the effect of increasing off-diagonal elements $\tilde{A}_{xy}$ causing a change in decay frequency. On the contrary, simulations using the proposed third order correction of the equilibrium function result in a significant difference. The effect stems from higher order excitations present in the macroscopic fields. Increasing $\tilde{A}_{xy}$ results in a notably decreasing decay frequency.  

%%%%%%%%%%%%%%%%%%%%%%%%%%%%%%%%%%%%%%%%%%%%%%%%%%%%%%%%%%%%%%%%%%%%%%%%%%%%%%%
\begin{figure}
    \centering
    \includegraphics[width=0.95\linewidth]{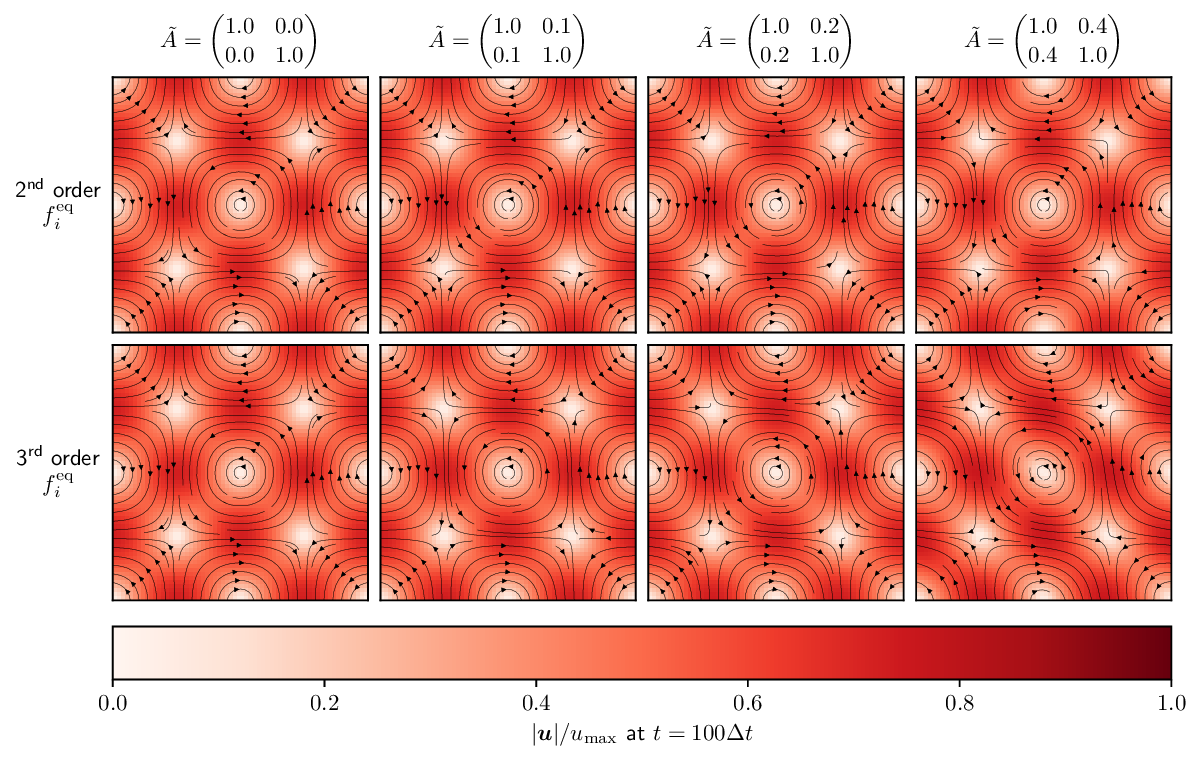}
    \caption{Velocity fields for Taylor-Green-vortex decay. Top row: second order equilibrium. Bottom row: third order equilibrium. From left to right: $\tilde{A}_{xy} = 0.0,\,0.1,\,0.2$, and $0.4$. Only with the third order the skewed nature of the velocity field becomes apparent.}
    \label{fig:TG_visual}
\end{figure}
%%%%%%%%%%%%%%%%%%%%%%%%%%%%%%%%%%%%%%%%%%%%%%%%%%%%%%%%%%%%%%%%%%%%%%%%%%%%%%%

Visually this result can be confirmed by the resulting flow patterns. \Cref{fig:TG_visual} shows the flow fields for each of the simulations at $t=100\Delta t$. Here, we can observe that only in the case of third order approximation of the equilibrium function the flow fields appear as skewed, while using second order equilibrium results in all cases in visually isotropic flow fields. From the form of {\Cref{eq:equil}} we expect a skewed field, because the introduction of $A$ is somewhat equivalent to an affine transformation of the coordinates.\\
Note, that in test cases without above mentioned higher order excitations, e.g. regular Poiseuille flow through a channel, both approximation orders would appear identical.

\section{Conclusion and Outlook}
\label{sec:conclusion}
We were able to derive a consistent lattice Boltzmann framework for anisotropic flow problems using an anisotropic equilibrium distribution. We provided the proof through Chapman-Enskog theory that the governing macroscopic equations are correctly derived as being an anisotropic version of the Navier-Stokes equations. Important properties of LBM, like locality of the collision operator or isotropy of the space and velocity lattices are retained.

Depending on the problem under consideration, an expansion of the equilibrium distribution to higher orders than usual might be necessary to capture the relevant phenomena, as we demonstrated with a Taylor-Green-vortex decay using an anisotropic BGK model.

An interesting application of this model in future work is a volume averaged \cite{Lautenschlaeger_2022,Marquardt_2024} composite collision \cite{Weinmiller_2026} where BGK and bounce-back collision are combined in an anisotropic scheme to model fluid flow in unresolved anisotropic porous media. Additionally, the modification to anisotropic advection-diffusion would come with interesting applications and is relatively straight-forward to perform. A related topic is the anisotropic equilibrium distribution itself and how it might be measured or computed. It will be worth investigating how known LBM forcing schemes can be incorporated into the anisotropic model and if modifications are necessary.

Other related questions that might be worth investigating in the future are the incorporation of external forces, anisotropic equilibrium distributions in more complex collision operators like MRT, or anisotropic multiphase flow.

% \paragraph*{Funding:}
% This work has partially received financial support from the \textit{European Union’s Horizon 2020 Research and Innovation Programme} within the project DEFACTO under grant number 875247.

\paragraph*{Acknowledgements:}
The presented research contributes to the Center for Electrochemical Energy Storage Ulm-Karlsruhe (CELEST).

\paragraph*{Author Contributions:}
Methodology, software, formal analysis, investigation, resources, data curation, visualization, writing (original draft): BK.
Conceptualization, validation: BK, JW.
Writing (Review\,\&\,Editing): BK, JW, AL, TD.
Supervision, project administration, funding acquisition: AL, TD.
All authors confirm that they have read the final manuscript and agree to its publication.

\paragraph*{Data availability:}
The data that support the findings of this study were generated by numerical simulations. The source code and parameters used to generate the simulations is publicly available. \cite{data}

\paragraph*{Conflicts of Interest:}
The authors declare no conflicts of interest.

\paragraph*{Abbreviations:}
We used the following abbreviations in the text.

\begin{tabular}{ll}
    BGK &   Bhatnagar–Gross–Krook \\
    LBM &   Lattice-Boltzmann method \\
    MRT &   Multi-Relaxation-Time
    
\end{tabular}

\printbibliography

\clearpage
%%%%%%%%%%%%%%%%%%%%%%%%
%      APPENDIX
%%%%%%%%%%%%%%%%%%%%%%%%
\appendix
\section*{Appendix}

\section{Useful Integral Identities}
\label{app:integrals}
The integrals in \Cref{sec:methods} can be solved with these helpful identities:
\begin{equation}
    \infint \exp\left(
        -\frac{(v_\ia - u_\ia)A_{\ia\ib}(v_\ib - u_\ib)}{2} 
    \right) \dd v 
    = \sqrt{\frac{(2\pi)^d}{\det A}},
\end{equation}
\begin{equation}
    \infint \exp\left(
        -\frac{(v_\ia - u_\ia)A_{\ia\ib}(v_\ib - u_\ib)}{2} 
    \right) v_\ic \dd v 
    = \sqrt{\frac{(2\pi)^d}{\det A}} u_\ic,
\end{equation}
\begin{equation}
    \infint \exp\left(
        -\frac{(v_\ia - u_\ia)A_{\ia\ib}(v_\ib - u_\ib)}{2} 
    \right) v_\ic v_\id \dd v 
    = \sqrt{\frac{(2\pi)^d}{\det A}} (A^{-1}_{\ic\id} + u_\ic u_\id),
\end{equation}
and finally
\begin{equation}
    \infint \exp\left(
        -\frac{(v_\ia - u_\ia)A_{\ia\ib}(v_\ib - u_\ib)}{2} 
    \right) v_\ic v_\id v_\ie \dd v 
    = \sqrt{\frac{(2\pi)^d}{\det A}} \left(
        A^{-1}_{\ic\id} u_\ie
        + A^{-1}_{\id\ie} u_\ic
        + A^{-1}_{\ie\ic} u_\id
        + u_\ic u_\id u_\ie
    \right).
\end{equation}

\section{Hermite polynomial expansion}
\label{app:Hermite}

Hermite polynomials of order $n$
\begin{equation}
    \Hn{n}(\v) = \frac{(-1)^n}{\omega(\v)}\nabla^{(n)}\omega(\v).
\label{eq:hermite_polynimals}
\end{equation}
are generated using the generating function%
%%%%%%%%%%%%%%%%%%%%%%%%%%%%%%%%%%%%%%%%%%%%%%%%%%%%%%%%%%%%%%%%%%%%%%%%%%%%%%%%%%%%%%%%%
\footnote{%
This function is not unique and could vary from what the reader might be familiar with. This variant generates the so-called \textit{probabilist's Hermite polynomials}.%
}
%%%%%%%%%%%%%%%%%%%%%%%%%%%%%%%%%%%%%%%%%%%%%%%%%%%%%%%%%%%%%%%%%%%%%%%%%%%%%%%%%%%%%%%%%
\begin{equation}
    \omega(\v) = \frac{1}{\sqrt{2\pi}^d} \exp \left( -\frac{v_\ia v_\ia}{2} \right).
\label{eq:generator}
\end{equation}

The object $\nabla^{(n)}$ is a tensor of rank $n$ which applies $n$ derivatives in the order of the indices, i.\,e.
\begin{equation}
    \nabla^{(n)} = \nabla_{\ia_1...\ia_n} 
    = \frac{\partial}{\partial v_{\ia_1}} ... \frac{\partial}{\partial v_{\ia_n}},
\end{equation}
which makes $\Hn{n}$ also a tensor of rank $n$.

With this, any function $g(\v)$ can be expanded in the basis of Hermite polynomial as
\begin{equation}
    g(\v) = \omega(\v) \sum_{n=0}^\infty \frac{1}{n!} 
    \langle \an{n}(g) , \Hn{n} \rangle.
\end{equation}
The expansion coefficients $\an{n}(g)$ are found by projecting $g(\v)$ onto the $n$th basis polynomial, i.e.\ 
\begin{equation}
    \an{n}(g) = \infint g(\v) \Hn{n}(\v) \dd v.
\label{eq:expansion_coefficients}
\end{equation}
The operation $\langle\cdot,\cdot\rangle$ is the full contraction of the tensors, i.\,e.
\begin{equation}
    \langle \an{n}(g) , \Hn{n} \rangle = \an{n}_{\ia_1...\ia_n}(g) \Hn{n}_{\ia_1...\ia_n} .
\end{equation}

From \Cref{eq:hermite_polynimals} we compute the first four Hermite polynomials
\begin{equation}
\begin{split}   
    &\Hn{0}(\v) = 1,\\
    &\Hn{1}(\v) = v_\ia,\\
    &\Hn{2}(\v) = v_\ia v_\ib - \delta_{\ia\ib},\\
    &\Hn{3}(\v) = v_\ia v_\ib v_\ic 
        - \delta_{\ia\ib} v_\ic
        - \delta_{\ib\ic} v_\ia
        - \delta_{\ic\ia} v_\ib.
\end{split}
\label{eq:hermite_polynomials}
\end{equation}
Using the integral identities in \Cref{app:integrals}, the expansion coefficients for $\feq$ as defined in \Cref{eq:equil} are
\begin{equation}
\begin{split}
    \an{0}(\feq) &= \rho, \\
    \an{1}_\ia(\feq) &= \rho u_\ia, \\
    \an{2}_{\ia\ib}(\feq) &= \rho (u_\ia u_\ib + K_{\ia\ib}), \\
    \an{3}_{\ia\ib\ic}(\feq) &= 
        \rho ( u_\ia u_\ib u_\ic
            + K_{\ia\ib} u_\ic
            + K_{\ib\ic} u_\ia
            + K_{\ic\ia} u_\ib ),
\end{split}
\end{equation}
where we define 
\begin{equation}
    K_{\ia\ib} = A^{-1}_{\ia\ib} - \delta_{\ia\ib}.
\end{equation}

\section{Gauss-Hermite Quadrature}
\label{app:quadrature}

Lets assume we need to evaluate an integral over the whole space in which the integrated function is a product of the Hermite generator $\omega$ and an arbitrary 1D polynomial $P^{(N)}(x)$ of order $N$. Such integrals can be approximated using a weighted finite sum of the polynomial, such that
\begin{equation}
    \infint \omega(x) P^{(N)}(x) \mathrm{d}x = \sum_{i=1}^n w_i P^{(N)}(x_i).
\end{equation}
The specific value of $n$ depends on the desired accuracy. In general $N \leq 2n-1$. A nice property of the Gauss-Hermite quadrature -- which sets it apart from most of the other quadrature variants -- is that it becomes exact for sufficiently large $n$, namely when $N=2n-1$ or $n=(N+1)/2$. The abscissae $x_i$ are the roots of the $n$th order Hermite polynomial and the weights can be computed from one order lower via
\begin{equation}
    w_i = \frac{n!}{[n H^{(n-1)}(x_i)]^2}.
\label{eq:weights}
\end{equation}
For example, a polynomial of order $N=5$ requires $n=3$ probing points. Hence, we need the third order Hermite polynomial $\Hn{3}$ to compute the abscissae and the second order polynomial $\Hn{2}$ for the weights. They can be derived from \Cref{eq:hermite_polynimals} by setting all the indices equal, such that 
\begin{equation}
    H^{(3)}(x) = x^3 - 3x,\quad H^{(2)}(x) = x^2 -1.
\end{equation}
The roots of $H^{(3)}$ are
\begin{equation}
    x_1 = 0, \quad x_{2,3} = \pm \sqrt{3},
\end{equation}
and using \Cref{eq:weights} yields the corresponding weights
\begin{equation}
    w_1 = \frac{2}{3}, \quad w_{2,3} = \frac{1}{6}.
\end{equation}

The extension to $d$ dimensions is straight-forward because of the two following properties of the generator and an arbitrary polynomial of order $N$, $P^{(N)}$:
\begin{subequations}
\begin{align}
    \omega(\x) = \prod_{\ia=1}^d \omega(x_\ia), \\
    P^{(N)}(\x) = \sum_{ \{\vec{k}\} } a_{\vec{k}} \prod_{\ia=1}^d x_\ia^{N_\ia},
\end{align}
\end{subequations}
 where the multi-index $\vec{k}$ runs over the set $\{\vec{k}\} = \{ (N_1,...,N_d) \in \mathbb{N}^d \mid \sum_\ia N_\ia \leq N \}$.

That way, the $d$-dimensional expression only includes 1D integrals, i.\,e.
\begin{equation}
    \infint \omega(\x) P^{(N)}(\x) \dd\x = \sum_{\{\vec{k}\}} a_{\vec{k}} \prod_{\ia=1}^d \infint \omega(x_\ia) x_\ia^{N_\ia} \mathrm{d}x_\ia,
    \label{eq:quadrature_general1}
\end{equation}
and the quadrature becomes
\begin{equation}
    \sum_{\{\vec{k}\}} a_{\vec{k}} \prod_{\ia=1}^d \infint \omega(x_\ia) x_\ia^{N_\ia} \mathrm{d}x_\ia
    = \sum_{\{\vec{k}\}} a_{\vec{k}} 
    \prod_{\ia=1}^d \sum_{i_\ia=1}^{n_\ia} w_{i_\ia} x_{i_\ia}^{N_\ia}.
\label{eq:quadrature_general2}
\end{equation}
With the assumption that all dimensions are expanded to the same degree, i.e.\ $n_\ia=n$, one can immediately construct the D2Q9 and D3Q27 lattices. Through symmetry arguments, cf.\ Appendix in Ref.\ \cite{Shan_2006}, one can reduce the lattice for $d=3$ to D3Q19 without loss of precision in the Navier-Stokes regime. By mapping the products of weights to a single running index, $w_{i1}...w_{id} \mapsto w_i$, one can rewrite \Cref{eq:quadrature_general2} with a single sum
\begin{equation}
    \infint \omega(\x) P^{(N)}(\x) \dd\x = \sum_{i=1}^q w_i P^{(N)}(\x_i)
\end{equation}
for the general lattice D$d$Q$q$.

\section{Series Expansion of the Discrete Boltzmann Equation}
\label{app:Taylor_expansion}
Let $g:\mathbb{R}^d \to \mathbb{R}$ an $N$-times continuously differentiable function. Then
\begin{equation}
    g(x_\ia + h_\ia) 
    = g(x_\ia) + \sum_{j=1}^N
        \frac{1}{j!} (h_\ib \partial_\ib)^j g(x_\ia) + R,
\end{equation}
where $R$ is a remainder of order $N+1$. Note that the $\ib$-term is summed over because of the repeated index such that it corresponds to a inner product.
With the definitions $g(x_\ia)=f_i(t,\x)$, $x_\ia = (t,\x)$, and $h_\ia=(\Delta t, \c_i\Delta t)$ we can write the discrete Boltzmann equation as
\begin{equation}
    f_i(t+\Delta t, \x+\c_i\Delta t) - f_i(t,\x) 
        = \sum_{j=1}^N \frac{\Delta t^j}{j!}(\partial_t + c_{i\ib}\partial_\ib)^j f_i(t,\x) = \Omega_i \Delta t.
\label{eq:Taylor_expansion}
\end{equation}

\section{Higher Order Moments of the Equilibrium}
\label{app:moments}
Although the physical observables are given by the zeroth and first order moments of the distribution function, the Chapman-Enskog analysis requires additionally the second and third order moments of the equilibrium function. For the continuous case, those can be computed using the integral identities in \Cref{app:integrals} and given by
\begin{equation}
    \mathcal{M}_{\ia\ib}(\feq) = \infint \feq(\rho,\u,\v) v_\ia v_\ib \dd{v}
    = \rho \left( 
          A^{-1}_{\ia\ib}
        + u_\ia u_\ib
    \right),
\end{equation}
and
\begin{equation}
    \mathcal{M}_{\ia\ib\ic}(\feq) = \infint \feq(\rho,\u,\v) v_\ia v_\ib v_\ic \dd{v}
    = \rho \left( 
          A^{-1}_{\ia\ib} u_\ic
        + A^{-1}_{\ib\ic} u_\ia
        + A^{-1}_{\ic\ia} u_\ib
        + u_\ia u_\ib u_\ic
    \right),
\end{equation}

In the discrete velocity space, we have
\begin{equation}
    M_{\ia\ib}(\feq_i) = \sum_i \feq_i v_{i\ia} v_{i\ib}
    = \rho \left( 
          A^{-1}_{\ia\ib}
        + u_\ia u_\ib
    \right),
\end{equation}
and
\begin{equation}
    M_{\ia\ib\ic}(\feq_i) = \sum_i \feq_i v_{i\ia} v_{i\ib} v_{i\ic}
    = \rho \left( 
          A^{-1}_{\ia\ib} u_\ic
        + A^{-1}_{\ib\ic} u_\ia
        + A^{-1}_{\ic\ia} u_\ib
        + u_\ia u_\ib u_\ic
    \right).
\end{equation}
Using the compatible velocity lattice, where $\v_i = \c_i / \cs$ and $\u = \w / \cs$, we find
\begin{equation}
    M_{\ia\ib}(\feq_i) =\rho \left( \cs^2 A^{-1}_{\ia\ib} + s_\ia s_\ib \right),
\end{equation}
and
\begin{equation}
M_{\ia\ib\ic}(\feq_i) 
    = \rho \left( 
          \cs^2 A^{-1}_{\ia\ib} s_\ic
        + \cs^2 A^{-1}_{\ib\ic} s_\ia
        + \cs^2 A^{-1}_{\ic\ia} s_\ib
        + s_\ia s_\ib s_\ic
    \right).
\label{eq:mom_diff1}
\end{equation}

The reader should note that the moments of a full and truncated equilibrium distributions will be identical up to the order of expansion. Since we expanded $\feq$ to third order, all the above moments remain identical. If we had expanded only to second order, the third moment of the second order equilibrium $\hat{\feq}$ would lack the anisotropic correction and would therefore read
\begin{equation}
\hat{M}_{\ia\ib\ic}(\hat{\feq}_i) 
    = \rho \left( 
          \cs^2 \delta_{\ia\ib} s_\ic
        + \cs^2 \delta_{\ib\ic} s_\ia
        + \cs^2 \delta_{\ic\ia} s_\ib
        + s_\ia s_\ib s_\ic
    \right).
\label{eq:mom_diff2}    
\end{equation}

\section{Navier-Stokes Equations for Single-relaxation Time Collisions}
\label{app:navier_stokes}

The force-free isotropic Navier-Stokes equations for a fluid of density $\rho$ and a macroscopic velocity $\u$ under pressure $p$ can be written as
\begin{equation}
    \partial_t(\rho u_\ia) 
    + \partial_\ib ( \rho u_\ia u_\ib ) 
    = 
    - \partial_\ia p
    + \partial_\ib \left[ 
        \eta
            \left(
                \partial_\ib u_\ia
                + \partial_\ia u_\ib
            \right)
        +
        \delta_{\ia\ib} 
        \left(\zeta - \frac{2}{3}\eta\right)
        \partial_\ic u_\ic
        \right].
\label{eq:appNS1}
\end{equation}
We introduced the shear viscosity $\eta$ and the bulk or volume viscosity $\zeta$. The Kronecker delta hints at the fact that the viscosities are actually not scalars but tensor quantities. Since we investigate the single-relaxation time (BGK) collision under isothermal assumption, the viscosities cannot be varied independently. The isentropic equation of state for pressure $p$ and density $\rho$ is
\begin{equation}
    p = p_0 \left(\frac{\rho}{\rho_0}\right)^\gamma,
\end{equation}
where we introduced the adiabatic index $\gamma$. In general, the viscosities are related through
\begin{equation}
    \zeta = \eta \left( \frac{5}{3} - \gamma \right).   
\end{equation}
The isothermal limit imposed by the equilibrium distribution \Cref{eq:equil}, however, is characterized by $\gamma = 1$, such that 
\begin{equation}
    \zeta = \frac{2}{3}\eta.  
\end{equation}
This relation sets the last term in \Cref{eq:appNS1} to zero, and the isothermal Navier-Stokes equations read
\begin{equation}
      \partial_t(\rho u_\ia) 
    + \partial_\ib ( \rho u_\ia u_\ib ) 
    = 
    - \partial_\ia p
    + \partial_\ib \left[ 
        \eta
            \left(
                \partial_\ib u_\ia
                + \partial_\ia u_\ib
            \right)
        \right].
% \label{eq:appNS2}  
\end{equation}

The shear viscosity was already identified as a tensor. But the single-relaxation time collision leads to the relations $\eta \propto \rho$ and $p \propto \rho$, and hence, $\eta \propto p$. We therefore assume that not only $\eta$ but also $p$ is implicitly a tensor that carries the anisotropy. We hence assume to arrive at Navier-Stokes equations where $p$ and $\eta$ both are anisotropic tensors, i.e.
\begin{equation}
      \partial_t(\rho u_\ia) 
    + \partial_\ib ( \rho u_\ia u_\ib ) 
    = 
    - \partial_\ib \left(C_{\ia\ib}p\right)
    + \partial_\ib \left[ 
        \eta
            \left(
                C_{\ib\ic}\partial_\ic u_\ia
                + C_{\ia\ic}\partial_\ic u_\ib
            \right)
        \right],
% \label{eq:appNS2}  
\end{equation}
where $C$ is a tensor related to $A$. The Chapman-Enskog analysis in \Cref{sec:CET} shows that this is indeed the case, and that $C \propto A^{-1}$. 

More details on the thermodynamic relations can be found in \cite{Krueger_2017:LBM_Book} and the cited references therein.

\section{Explicit Expression for the Perturbation Moment}
\label{app:perturbation_moment}

During the Chapman-Enskog analysis we encounter the equation
\begin{equation}
    \partial_t(\rho u_\ia) + \partial_\ib M_{\ia\ib} (\feq_i) = -\epsilon 
    \left( 1 - \frac{\Delta t}{2\tau} \right) 
    \partial_\ib M_{\ia\ib}(f^{(1)}_i).
\label{eq:CE3b}
\end{equation}
and
\begin{equation}
    \epsilon \partial_t^{(1)} M_{\ia\ib}(\feq_i) 
    + \epsilon \partial_\ic^{(1)} M_{\ia\ib\ic}(\feq_i)
    = -\frac{\epsilon}{\tau}  M_{\ia\ib}(f_i^{(1)}),
\label{eq:CE4b}
\end{equation}
which we want to compute here as an explicit expression of the observables $\rho$ and $\u=\vec{s}/\cs$. The steps are provided in Ref.\ \cite{Krueger_2017:LBM_Book} and require only slight modifications.

Using the expression we found in \Cref{app:moments}, we write out the time derivative of the second order equilibrium moment, i.e.\ 
\begin{equation}
    \partial_t^{(1)} M_{\ia\ib}(\feq_i) 
    = \partial_t^{(1)} \left(
         \cs^2 A^{-1}_{\ia\ib} \rho
        + s_\ia s_\ib \rho
    \right).
\label{eq:F1}
\end{equation}
The second term on the right-hand side needs to be rewritten using a variation of the product rule. For some functions $f$, $g$, and $h$ with a generic derivative of $f$ denoted as $f'$, we can write
\begin{equation}
    (fgh)' = f(gh)' + ghf',
\end{equation}
and since
\begin{equation}
    hf' = (fh)' - fh',
\end{equation}
we can write the first expression as
\begin{equation}
    (fgh)' = f(gh)' + g(fh)' - fgh'.
\label{eq:product_rule}
\end{equation}
Applying the rule to \Cref{eq:F1}, we find
\begin{equation}
    \partial_t^{(1)} M_{\ia\ib}(\feq_i) 
    = \cs^2 A^{-1}_{\ia\ib} \partial_t^{(1)} \rho
        + s_\ia \partial_t^{(1)} (s_\ib \rho)
        + s_\ib \partial_t^{(1)} (s_\ia \rho)
        - s_\ia s_\ib \partial_t^{(1)} \rho
\label{eq:F2}
\end{equation}
Next, we rewrite the first order balance equations for mass and momentum, \Cref{eq:1st_order_mass,eq:1st_order_momentum} as
\begin{equation}
    \partial_t^{(1)} \rho 
    = - \partial_\ia^{(1)}(s_\ia \rho),
\end{equation}
and
\begin{equation}
    \partial_t^{(1)}(s_\ia \rho)
    = - \partial_\ib^{(1)} \left(
         \cs^2 A^{-1}_{\ia\ib} \rho
        + s_\ia s_\ib \rho
    \right),
\end{equation}
and plug them into \Cref{eq:F2} to eliminate some time derivatives, i.e.\ 
\begin{align}
\begin{split}
    \partial_t^{(1)} M_{\ia\ib}(\feq_i) 
    = &- \cs^2 A^{-1}_{\ia\ib} \partial_\ic^{(1)}(s_\ic \rho)
        - s_\ia \partial_\ic^{(1)} \left(
              \cs^2 A^{-1}_{\ib\ic} \rho
            + s_\ib s_\ic \rho
        \right)
        \\
        &- s_\ib \partial_\ic^{(1)} \left(
              \cs^2 A^{-1}_{\ia\ic} \rho
            + s_\ia s_\ic \rho
        \right) 
        + s_\ia s_\ib \partial_\ic^{(1)}(s_\ic \rho).
\end{split}
\label{eq:F3}
\end{align}
This can be rewritten as
\begin{align}
\begin{split}
    \partial_t^{(1)} M_{\ia\ib}(\feq_i) 
    = &- \cs^2 A^{-1}_{\ia\ib} \partial_\ic^{(1)}(s_\ic \rho)
        - s_\ia \cs^2 A^{-1}_{\ib\ic} \partial_\ic^{(1)} \rho
        - s_\ib \cs^2 A^{-1}_{\ia\ic} \partial_\ic^{(1)} \rho
        \\
        &+ s_\ia s_\ib \partial_\ic^{(1)}(s_\ic \rho)
        - s_\ia \partial_\ic^{(1)} \left(
            s_\ib s_\ic \rho
        \right)
        - s_\ib \partial_\ic^{(1)} \left(
            s_\ia s_\ic \rho
        \right) ,
\end{split}
\end{align}
to apply the product rule in \Cref{eq:product_rule} in reverse to the second line, such that
\begin{equation}
    \partial_t^{(1)} M_{\ia\ib}(\feq_i) 
    = - \cs^2 A^{-1}_{\ia\ib} \partial_\ic^{(1)}(s_\ic \rho)
      - \cs^2 A^{-1}_{\ib\ic} s_\ia \partial_\ic^{(1)} \rho
      - \cs^2 A^{-1}_{\ia\ic} s_\ib \partial_\ic^{(1)} \rho
        -\partial_\ic^{(1)} (s_\ia s_\ib s_\ic \rho).
\end{equation}

Fortunately, the expression for the third order moment requires less effort and can be computed directly as
\begin{equation}
    \partial_\ic^{(1)} M_{\ia\ib\ic}(\feq_i)
    = \partial_\ic^{(1)} \left(
          \cs^2 A^{-1}_{\ia\ib} s_\ic \rho
        + \cs^2 A^{-1}_{\ib\ic} s_\ia \rho
        + \cs^2 A^{-1}_{\ia\ic} s_\ib \rho
        + s_\ia s_\ib s_\ic \rho
    \right).
\end{equation}

Going back to \Cref{eq:CE4b} we now found
\begin{align}
\begin{split}
    M_{\ia\ib}(f_i^{(1)}) = 
      \tau \cs^2 A^{-1}_{\ia\ib} \partial_\ic^{(1)}(s_\ic \rho)
    + &\tau \cs^2 A^{-1}_{\ib\ic} s_\ia \partial_\ic^{(1)} \rho
    +  \tau \cs^2 A^{-1}_{\ia\ic} s_\ib \partial_\ic^{(1)} \rho
    + \tau \partial_\ic^{(1)} (s_\ia s_\ib s_\ic \rho)
    \\
    - & \tau \partial_\ic^{(1)} \left(
          \cs^2 A^{-1}_{\ia\ib} s_\ic \rho
        + \cs^2 A^{-1}_{\ib\ic} s_\ia \rho
        + \cs^2 A^{-1}_{\ia\ic} s_\ib \rho
        + s_\ia s_\ib s_\ic \rho
    \right),
\end{split}
\end{align}
which, using the regular product rule, simplifies to
\begin{align}
\begin{split}
    M_{\ia\ib}(f_i^{(1)}) 
    &= 
    - \tau \left[ 
          \cs^2 A^{-1}_{\ib\ic} \left(
            \partial_\ic^{(1)} (s_\ia \rho)
            - s_\ia \partial_\ic^{(1)} \rho
        \right)
        + \cs^2 A^{-1}_{\ia\ic} \left( 
            \partial_\ic^{(1)} (s_\ib \rho)
            - s_\ib \partial_\ic^{(1)} \rho
        \right)
    \right] 
    \\
    &= 
    - \tau \rho \cs^2 \left[ 
          A^{-1}_{\ib\ic} \partial_\ic^{(1)} s_\ia
        + A^{-1}_{\ia\ic} \partial_\ic^{(1)} s_\ib
    \right].
\end{split}
\end{align}
N.\,B.: This final expression is only that simple if we expand the equilibrium to the third order. Only then $M_{\ia\ib\ic}$ also carries the $A^{-1}$-terms, cf.\ \Cref{app:moments} 
and compare \Cref{eq:mom_diff1} with \Cref{eq:mom_diff2}. With a second order equilibrium the mixed order sum of equilibrium moments in \Cref{eq:CE4b} will include $K$-terms (\Cref{eq:K-term}) as discretization errors. Additionally, we would see a $\mathcal{O}(u^3)$ discretization error, which is also present in second order isotropic LBM and usually neglected.

\end{document}